\documentclass[aps,prb,superscriptaddress,twocolumn]{revtex4}

\usepackage{amsmath,amssymb}
\usepackage{graphicx}
\usepackage{feynmf}
\usepackage{color}
\usepackage{dsfont}

\newcommand{\rs}{\rm \scriptscriptstyle}

\begin{document}

\title{Extended quantum Maxwell demon acting over macroscopic distances}

\date{\today}

\author{A.V.\ Lebedev}
\affiliation{Moscow Institute of Physics and Technology, Institutskii per.\ 9, Dolgoprudny,
141700, Moscow District, Russia}
\affiliation{Theoretische Physik, Wolfgang-Pauli-Strasse 27, ETH Zurich,
CH-8093 Z\"urich, Switzerland}

\author{G.B.\ Lesovik}
\affiliation{Moscow Institute of Physics and Technology, Institutskii per.\ 9, Dolgoprudny,
141700, Moscow District, Russia}

\author{V.M.\ Vinokur}
\affiliation{Materials Science Division, Argonne National Laboratory,
		9700 S. Cass Avenue, Argonne, IL 60439, USA}

\author{G.\ Blatter}
\affiliation{Theoretische Physik, Wolfgang-Pauli-Strasse 27, ETH Zurich,
CH-8093 Z\"urich, Switzerland}

\begin{abstract}
A quantum Maxwell demon is a device that can lower the entropy of a quantum
system by providing it with purity. The functionality of such a quantum demon
is rooted in a quantum mechanical SWAP operation exchanging mixed and pure
states.  We describe the setup and performance of a quantum Maxwell demon that
purifies an energy-isolated system from a distance. Our cQED-based design
involves two transmon qubits, where the mixed-state target qubit is purified
by a pure-state demon qubit connected via an off-resonant transmission line;
this configuration naturally generates an iSWAP gate. Although less powerful
than a full SWAP gate, we show that assuming present-day performance
characteristics of a cQED implementation, such an extended quantum Maxwell
demon can purify the target qubit over macroscopic distances on the order of
meters and tolerates elevated temperatures of the order of a few Kelvin in the
transmission line.
\end{abstract}

\maketitle

\section{Introduction}\label{sec:intro}

Maxwell's demon \cite{Maxwell} is a putative device that is capable of
observing and controlling the microscopic degrees of freedom of a
thermodynamic system. The existence of such a demon permits the cyclic
extraction of work in a heat engine with unit efficiency and thus apparently
violates the Second Law of Thermodynamics.  After a century long debate
\cite{Maruyama:2009}, it has been realized by Landauer \cite{Landauer:1961}
and by Bennett \cite{Bennett:2003} that the demon's functionality requires a
memory in which to store the results of its observations. The cyclic operation
of the engine then must include an element that erases the information in the
demon's memory.  According to Landauer, this erasure involves an entropy
increase per bit of $\Delta S = k_{\rs B}\ln{2}$. A crucial element in
furthering the argument is to include the demon, which is situated in the
immediate proximity of the system, as a part of the system.  As a natural
consequence, the thermodynamic cost of erasing the demon's memory then is
accounted for in the engine's overall entropy budget, thereby restoring the
validity of the Second Law.  Thermodynamic machines utilizing the
functionality of such a locally operating classical Maxwell demon have been
recently demonstrated in several systems \cite{Toyabe:2010, Koski:2014,
Koski:2014_2, Roldan:2014}.

Within a quantum mechanical framework, new opportunities arise, e.g., a demon
has been conceived \cite{lesovik:2016} that allows to reduce the entropy of an
energy-isolated system. This has inspired the proposal for an engine that
features separated cycles \cite{lebedev:2016,Kirsanov:2018}, an energy cycle
that transforms heat into work without thermal waste and an entropy cycle that
restores the Second Law. These findings motivate the question about the
distance over which such a quantum Maxwell demon can perform its beneficial
action.  In this paper, we analyze the performance of an extended quantum
Maxwell demon (QMD); specifically, we determine the demon's maximal spatial
separation and its operating conditions that allow for an entropy reduction of
a distant energy-isolated quantum system. This lifts the question about a
possible local violation of the Second Law to a quantitative level.
Furthermore, such separation between the system and the demon is of practical
relevance as it naturally protects the system against undesired heating during
the demon's Landauer purification; within the context of quantum information
processing, an extended demon can be used to feed pure states to an ongoing
quantum computation.  In a wider context, the coherent communication between
quantum systems separated by large distances \cite{Siddiqi:2014,Schuster:2018}
is of great relevance, e.g., in distant entanglement \cite{Devoret:2018}, in
recent Bell tests \cite{Hanson:2015,Wallraff:2018} and in quantum state
transfer \cite{Schoelkopf:2018}.
\begin{figure}[ht]
   \includegraphics[width=8.6cm]{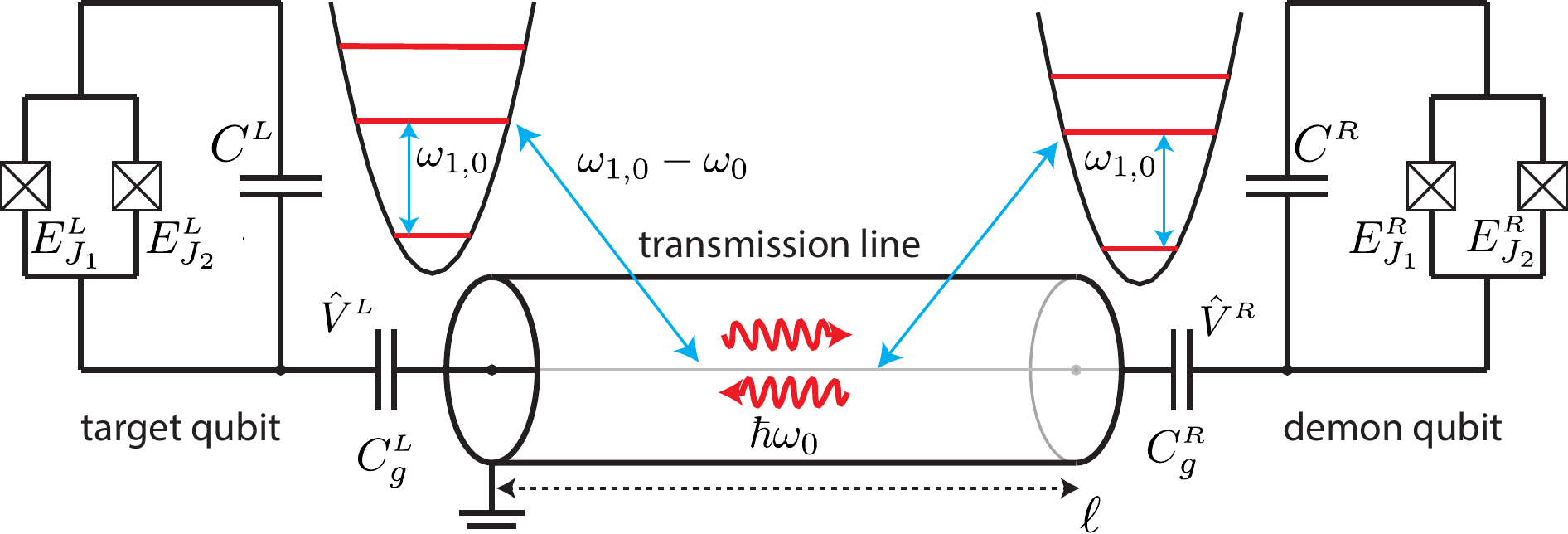}
        \caption{\label{fig:setup}
Schematic setup for an extended quantum Maxwell demon swapping the mixed state
of the target qubit (left) with the pure state of the demon qubit (right) via
an autonomous and energy-conserving process. The qubits are formed by the two
low-energy states of a transmon device comprised of a SQUID loop with two
Josephson junctions with energies $E_{J_1}^\alpha$ and $E_{J_2}^\alpha$,
capacitively shunted by a large capacitor $C^\alpha$ ($\alpha = L,R$); for the
transmon, $E_C^\alpha = e^2/2C^\alpha \ll E_{J_{1,2}}^\alpha$.  The
off-resonant transmission line of length $\ell$ connecting the two qubits is
filtered to a frequency band $\Delta\omega$ around $\omega_0$ and generates an
$XY$-type interaction that remains effective at macroscopic distances $\ell$
and elevated temperature $T_{\rm line} \gg T_{\rm qubit}$.
}
\end{figure}

A first version of a quantum Maxwell demon has been proposed by Lloyd
\cite{Lloyd:1997} in the context of nuclear magnetic resonance experiments,
see also Refs.\ [\onlinecite{Parrondo:2011, Kim:2011, Uzdin:2015}], based on
the idea that such a device exchanges the mixed quantum state of a {\it
target} system with a more pure quantum state of the {\it demon}. Such an
exchange is realized in the course of the coherent unitary evolution of the
joint target--demon system. In contrast to the classical version, the quantum
demon utilizes its quantum purity as a thermodynamic resource and does not
measure the state of the target system, hence its functionality roots in
purity rather than information \cite{Esposito:2014}.  Several proposals for
QMD-assisted thermodynamic machines have been suggested \cite{Quan:2006,
Wang:2009, Brask:2015, Pekola:2016, Hofer:2016, lebedev:2016} but only few
have been realized experimentally \cite{Vidrighin:2016, Camati:2016,
Cottet:2017}. Here, we describe a practical design of a spatially distributed
QMD setup that is able to purify the state of a distant quantum system, the
target qubit, by deterministically transforming its unknown mixed state into a
given pure quantum state that is provided by the demon.

Below, we focus on a circuit QED implementation with two distant transmon
qubits \cite{devoret:2007}, the target- and the demon-qubits, that are
capacitively coupled via a bosonic bath in the form of a transmission line
\cite{pozar:2012}, see Fig.\ \ref{fig:setup}. This setup leads to an XY-type
interaction between the qubits that naturally generates an iSWAP gate---the
latter's purification power is reduced as compared to a full SWAP gate.
However, the simpler implementation and enhanced robustness to decoherence of
the iSWAP gate motivates us to focus on this simpler version of a quantum
demon in our analysis below. We then address two main questions: i) what
spatial separation between target and demon qubits can be achieved for such an
extended quantum demon, given the finite coherence time of the components, and
ii) what are the requirements for the thermodynamic state of the bosonic bath,
the transmission line, that mediates the interaction between the systems. We
find that, given typical parameters describing present days cQED systems, a
distance $\ell$ of the order of a few meters can be reached with a
transmission line operating in the Kelvin range, i.e., about two orders of
magnitude higher than the operating temperature of the qubits.

\section{cQED setup of extended quantum demon}\label{sec:demon}

A target--demon setup of the type outlined above is described by the
Hamiltonian
\begin{eqnarray}\label{eq:ham}
      &&\hat{H} = \sum_{\alpha = {\rm \scriptscriptstyle L,R}} \sum_{i=0}^\infty
	\epsilon_i^{\alpha} \, |i\rangle_{\!\alpha} \,{}_\alpha\!\langle i|
       + \hat{H}_\mathrm{line}
      \\
      &&+ \sum_{\alpha = {\rm \scriptscriptstyle L,R}} \sum_{i=0}^\infty
	 \bigl[ q_{i+1,i}^{\alpha} \hat{V}^{\alpha}(x^\alpha) \>
         |i\!+\!1\rangle_{\!\alpha}\, {}_\alpha\! \langle i| + {\rm h.c.}\bigr],
      \nonumber
\end{eqnarray}
where $\epsilon_i^{\alpha}$ and $|i\rangle_\alpha$ describe the energy levels
of the left and right ($\alpha = {\rm L,R}$) transmon qubits positioned at
$x^{\rm \scriptscriptstyle L} = -\ell/2$ and $x^{\rm \scriptscriptstyle R} =
+\ell/2$ and $\hat{V}(x)$ is the voltage at position $x$ along the
transmission line. The latter generates an additional voltage drop
$\beta^\alpha \hat{V}^{\alpha}(x^\alpha)$ across the qubit capacitor, where
the reduction factors $\beta^\alpha = C^\alpha_\mathrm{g}/(C^\alpha_\mathrm{g}
+ C^\alpha)$ account for the capacitors' geometries, see Fig.\
\ref{fig:setup}. This voltage drop couples to the Cooper pairs
$\hat{n}^{\alpha}$ transferred between the transmon capacitor with the
effective charge $q_{i+1,i}^{\alpha} = 2e \, \beta^\alpha {}_\alpha\!\langle
i+1|\hat{n}^{\alpha}|i\rangle_{\!\alpha}$, where we incorporate the
geometrical factor $\beta^\alpha$.  Finally, the Hamiltonian of the
transmission line is
\begin{equation}
       \hat{H}_\mathrm{line} = \frac12  \int dx\, \bigl\{ {\cal C}
	[\hat{V}(x)]^2 + {\cal L} [\hat{I}(x)]^2\bigr\},
\end{equation}
where $\hat{I}(x)$ and $\hat{V}(x)$ are electric current- and voltage-fields
along the transmission line with ${\cal C}$ and ${\cal L}$ the capacitance and
inductance per unit length. The fields $\hat{I}(x)$ and $\hat{V}(x)$ are
obtained through a standard canonical quantization procedure of the
transmission line equations \cite{She:1965}, see Appendix \ref{app:Trans_Line}.

In order to allow for optimizing the performance of the device (see below), we
assume the modes $\omega_k$ of the transmission line to be off-resonant with
respect to the transition frequencies $\omega_{i,j}^{\alpha} =
(\epsilon_i^{\alpha} - \epsilon_j^{\alpha})/\hbar$ of the transmons. The
ensuing weak coupling allows for a perturbative treatment of the qubit--mode
interaction. We make use of a unitary transformation of \eqref{eq:ham},
$\hat{H} \to \hat{\cal H} = \hat{U}\hat{H} \hat{U}^\dagger$, in order to
eliminate the transmission-line modes to lowest order. We choose the ansatz
$\hat{U} = \exp\bigl[ \hat{S} - \hat{S}^\dagger \bigr]$ and $\hat{S} =
\sum_{\alpha,i} q_{i+1,i}^{\alpha} |i+1\rangle_{\!\alpha}
\,{}_{\alpha\!}\langle i|\, \hat{Q}^\alpha_i$, where $\hat{Q}^\alpha_i$ is a
linear form of the bosonic operators, see Appendix \ref{app:Int_Ham} for
details. A rotating wave approximation then provides us with an effective
interaction between the qubits mediated via virtual-photon exchange
\cite{Majer:2007},
\begin{equation}
      \hat{H}_\mathrm{int} = \sum_{ij} J_{ij} |i+1\rangle_{
   \rm \scriptscriptstyle L}\, {}_{\rm\scriptscriptstyle L\!}
      \langle i| \otimes |j\rangle_{\rm \scriptscriptstyle R}\,
    {}_{\rm\scriptscriptstyle R\!} \langle j\!+\!1| + \mathrm{h.c.}
\end{equation}
The effective couplings $J_{ij}$ involve the commutator
$\bigl[\hat{Q}^\alpha_i, [\hat{V}^\beta]^\dagger \bigr]$ between photonic
field operators at the opposite ends of the transmission line that contributes
the factor (with $\alpha = {\rm R, L}$ and $\alpha \neq \beta$; we assume a
long transmission line $\ell \gg \lambda$, where $\lambda$ is the wavelength
of the transmission line modes)
\begin{equation}
   \bigl[\hat{Q}^\alpha_i, [\hat{V}^\beta]^\dagger \bigr]
    = \frac1{2 {\cal C} \ell} \int d\omega \, \frac{\omega_{i+1,i}^{\alpha}}
    {(\omega_{i+1,i}^{\alpha}-\omega)^2},
\end{equation}
and its combination with the charge factors $q_{i+1,i}^{\alpha}$ provides us
with the expression for the effective couplings
\begin{equation}
      J_{ij} \!=\! \frac{q_{i+1,i}^{\rm \scriptscriptstyle L}
	q_{j,j+1}^{\rm \scriptscriptstyle R}}{2{\cal C} \ell}
	\!\!\!\int\!\! d\omega \biggl[ \frac{\omega_{i+1,i}^{\rm \scriptscriptstyle
	L}}{(\omega_{i+1,i}^{\rm \scriptscriptstyle L} \!-\! \omega)^2}
	+ \frac{\omega_{j+1,j}^{\rm \scriptscriptstyle R}}
	{(\omega_{j+1,j}^{\rm \scriptscriptstyle R} \!-\! \omega)^2} \biggr].
      \label{eq:Jij}
\end{equation}
Its inverse-length dependence $J_{ij} \propto 1/\ell$ derives from the finite
propagation velocity $v=1/\sqrt{\cal LC}$ of the electromagnetic modes inside
the transmission line, implying an exchange time $\tau = \ell/v$ for the
virtual photons that scales linearly with distance $\ell$, thus reducing the
coupling strength between the qubits as they are further separated. In the
following, we assume that the transmission line modes are filtered to a
frequency interval $[\omega_0 - \Delta\omega/2, \omega_0 + \Delta\omega/2]$,
with $|\omega_0 - \omega_{i,j}^{\alpha}| \sim \omega_0$ and $\Delta\omega \ll
\omega_0$, simplifying \eqref{eq:Jij} to
\begin{equation}
      J_{ij} = \frac{q_{i+1,i}^{\rm \scriptscriptstyle L}
	q_{j,j+1}^{\rm \scriptscriptstyle R}}{2{\cal C} \ell} \!
	\biggl[ \frac{\Delta\omega\, \omega_{i+1,i}^{\rm \scriptscriptstyle
	L}}{(\omega_{i+1,i}^{\rm \scriptscriptstyle L} \!-\! \omega_0)^2}
	\!+\! \frac{\Delta\omega\, \omega_{j+1,j}^{\rm \scriptscriptstyle R}}
	{(\omega_{j+1,j}^{\rm \scriptscriptstyle R} \!-\! \omega_0)^2} \biggr].
\end{equation}
Furthermore, the transition energies of the target and demon qubits are chosen
equal, $\omega_{1,0}^{\rm \scriptscriptstyle L} = \omega_{1,0}^{\rm
\scriptscriptstyle R} = \omega_{1,0}$; otherwise, the interaction Hamiltonian
$H_\mathrm{int}$ would not conserve energy, implying that the transition
amplitudes $|i+1,j\rangle \to |i,j+1\rangle$ are suppressed due to oscillating
phase factors (this feature can be used to switch the coupling on/off). With
all other transitions chosen off-resonance, we can restrict the Hilbert space
of the two-qubit system to the two lowest pairs of energy states and arrive at
the effective system Hamiltonian
\begin{equation}
   \hat{H}_\mathrm{qb} \!=\!\!\!\! \sum_{\alpha= {\rm \scriptscriptstyle L,R}}\!\!
   \hbar\omega_{1,0}
   |1\rangle_{\!\alpha}\, {}_{\alpha\!} \langle 1| + J \bigl[ |1,0\rangle \langle 0,1|
   \!+\! |0,1\rangle \langle 1,0| \bigr],
   \label{eq:qubitH}
\end{equation}
with a real-valued coupling constant $J\equiv J_{00}$,
\begin{equation}
   J = \kappa^{\rm \scriptscriptstyle L} \kappa^{\rm \scriptscriptstyle R}
   \frac{\Delta\omega\, \omega_{1,0}}
   {(\omega_{1,0}\!-\!\omega_0)^2}  \, \frac{hv}{\ell}.
   \label{eq:coupling}
\end{equation}
The dimensionless qubit--transmission-line couplings $\kappa^{\rm
\scriptscriptstyle L}$ and $\kappa^{\rm \scriptscriptstyle R}$ read (we use
$q_{j,j+1}^\alpha = -2ie \beta^\alpha [(1+j)/2]^{1/2}$ $(E_J^\alpha
/8E_C^\alpha)^{1/4}$)
\begin{equation}
      \kappa^\alpha = \beta^\alpha \Bigl(\frac{E_J^\alpha}
	{2E_C^\alpha}\Bigr)^{1/4} \sqrt{\frac{Z_0}{R_Q}},
\end{equation}
with $E_J^\alpha$ and $E_C^\alpha$ the Josephson- and charge energies of the
transmon qubit $\alpha$, $R_\mathrm{Q} = h/e^2$ is the resistance quantum, and
$Z_0 = 1/v{\cal C}$ is the characteristic impedance of the transmission line.
For typical values $\beta \sim 0.1$, $Z_0 = 50~\Omega$, and $E_J/E_C \sim
100$ one arrives at $\kappa \sim 0.01$.

\section{Demon (i)SWAP operation}\label{sec:SWAP}

\begin{figure}[ht]
   \includegraphics[width=8.4cm]{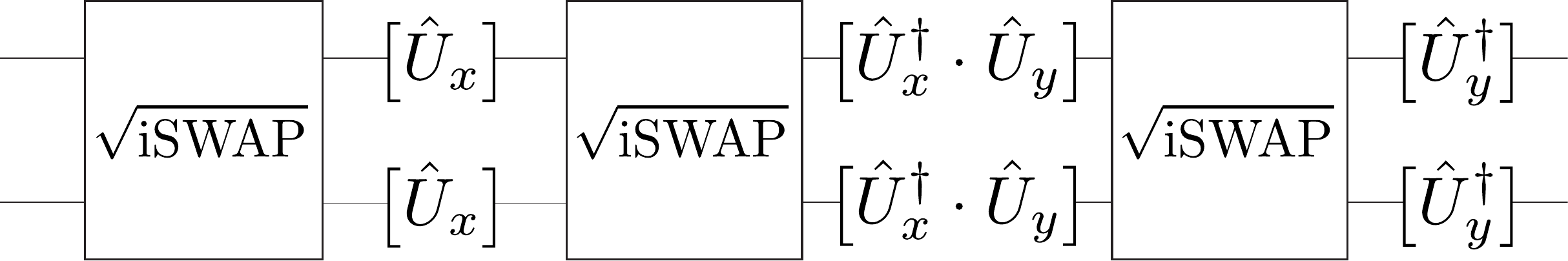}
        \caption{\label{fig:SWAP-iSWAP}
	Construction of the SWAP gate from three $\sqrt{\mathrm{iSWAP}}$ gates
	augmented with single-qubit rotations $\hat{U}_x (\pi/2)$ and
	$\hat{U}_y(\pi/2)$ and its conjugates.
}
\end{figure}

The XY-type interaction $H_\mathrm{XY} = (J/2) \bigl[ \hat\sigma_x
\hat\sigma_x + \hat\sigma_y \hat\sigma_y \bigr]$ in the two-qubit Hamiltonian
(\ref{eq:qubitH}) naturally leads to an iSWAP quantum gate \cite{Schuch:2003}
when running the evolution (we define $\omega_J = J/\hbar$)
\begin{equation}\label{eq:unitary}
      \hat{U}(\tau) = \left( \begin{array}{cccc}
      1&0&0&0\\
      0&\cos(\omega_J\tau)&-i\sin(\omega_J\tau)&0\\
      0&-i\sin(\omega_J\tau)&\cos(\omega_J\tau)&0\\
      0&0&0&1
      \end{array}\right)
\end{equation}
during the time $\tau_\mathrm{i{\scriptscriptstyle SWAP}} = \pi/2\omega_J$. On
the other hand, the optimal interaction for the SWAP gate $\mathrm{SWAP}
(\,|\psi\rangle_{\rm \scriptscriptstyle L} \,
|\chi\rangle_{\rm\scriptscriptstyle R}\,) = |\chi\rangle_{\rm \scriptscriptstyle
L}|\, \psi\rangle_{\rm \scriptscriptstyle R}$ is the isotropic Heisenberg
interaction $\hat{H}_\mathrm{XYZ} = (J/2)\bigl[ \hat\sigma_x \hat\sigma_x +
\hat\sigma_y \hat\sigma_y + \hat\sigma_z\hat\sigma_z\bigr]$; acting during the
time interval $\tau = h/4J$ it produces the unitary
\begin{equation}
   \hat{U}_\mathrm{SWAP} = e^{\frac{i\pi}4}\exp\bigl[ -i(\pi/4)
   \bigl( \hat\sigma_x\hat\sigma_x + \hat\sigma_y\hat\sigma_y + \hat\sigma_z
   \hat\sigma_z\bigr)\bigr].
       \label{eq:naturalSWAP}
\end{equation}
Given our setup, we have only the XY-interaction at our disposal, from which
one can generate a SWAP gate through a gate sequence involving square-roots of
iSWAP operations and single-qubit rotations, see Fig.\ \ref{fig:SWAP-iSWAP},
\begin{eqnarray}
      \label{eq:XYSWAP}
      &&\hat{U}_\mathrm{SWAP} = e^{\frac{i\pi}4}\bigl[\hat{U}_y^\dagger\otimes
      \hat{U}_y^\dagger e^{-i\frac\pi 8[ \hat\sigma_x\hat\sigma_x
    + \hat\sigma_y\hat\sigma_y]} \hat{U}_y\otimes \hat{U}_y\bigr]
      \\
      &&\bigl[\hat{U}_x^\dagger\otimes \hat{U}_x^\dagger e^{-i\frac\pi 8
      [ \hat\sigma_x\hat\sigma_x + \hat\sigma_y\hat\sigma_y]} \hat{U}_x\otimes
      \hat{U}_x\bigr]
      e^{-\frac{i\pi}8 [ \hat\sigma_x\hat\sigma_x + \hat\sigma_y\hat\sigma_y]},\quad
      \nonumber
\end{eqnarray}
where $\hat{U}_x = e^{-i\pi \hat\sigma_x/4}$ and $\hat{U}_y =
e^{-i\pi\hat\sigma_y/4}$ are $\pi/2$ rotations around the $x$- and $y$-axis,
respectively. Indeed, making use of the commutativity
$[\hat\sigma_\alpha\otimes \hat\sigma_\alpha, \hat\sigma_\beta \otimes
\hat\sigma_\beta] = 0$ for $\alpha,\beta = \{x,y,z\}$ and the unitary
transformations of the Pauli matrixes, $\hat{U}_x^\dagger \hat\sigma_y
\hat{U}_x = -\hat\sigma_z$ and $\hat{U}_y^\dagger \hat\sigma_x \hat{U}_y =
\hat\sigma_z$, one easily verifies the validity of Eq.~(\ref{eq:naturalSWAP}).
This SWAP gate implementation is twice faster than the one with three iSWAP
gates suggested in Ref.\ [\onlinecite{Schuch:2003}] and takes a time $\tau
= 1.5\, \tau_\mathrm{i{\scriptscriptstyle SWAP}}$ (we assume that all
single-qubit rotations can be done infinitely fast); according to the
discussion in Refs.\ [\onlinecite{Khaneja:2001,Vidal:2002}] it is optimal.

The SWAP gate can fully purify any state $\hat\rho_\mathrm{t}$ of the left
(target) qubit by exchanging its state with a pure state $\hat\rho_\mathrm{d}
= |\chi\rangle \langle \chi|$ of the right (demon) qubit. Moreover, preparing
the demon state with equal energy as the target qubit, $\mbox{Tr}
\{\hat{H}^{\rm \scriptscriptstyle R} \hat\rho_\mathrm{d}\} = \mbox{Tr}
\{\hat{H}^{\rm \scriptscriptstyle L} \hat\rho_\mathrm{t}\}$, one arrives at a
device which non-locally pushes the entropy of the target qubit to zero
without changing its energy, thus defining our desired quantum Maxwell demon.

\begin{figure}[h!]
\centering
\includegraphics[width=6cm]{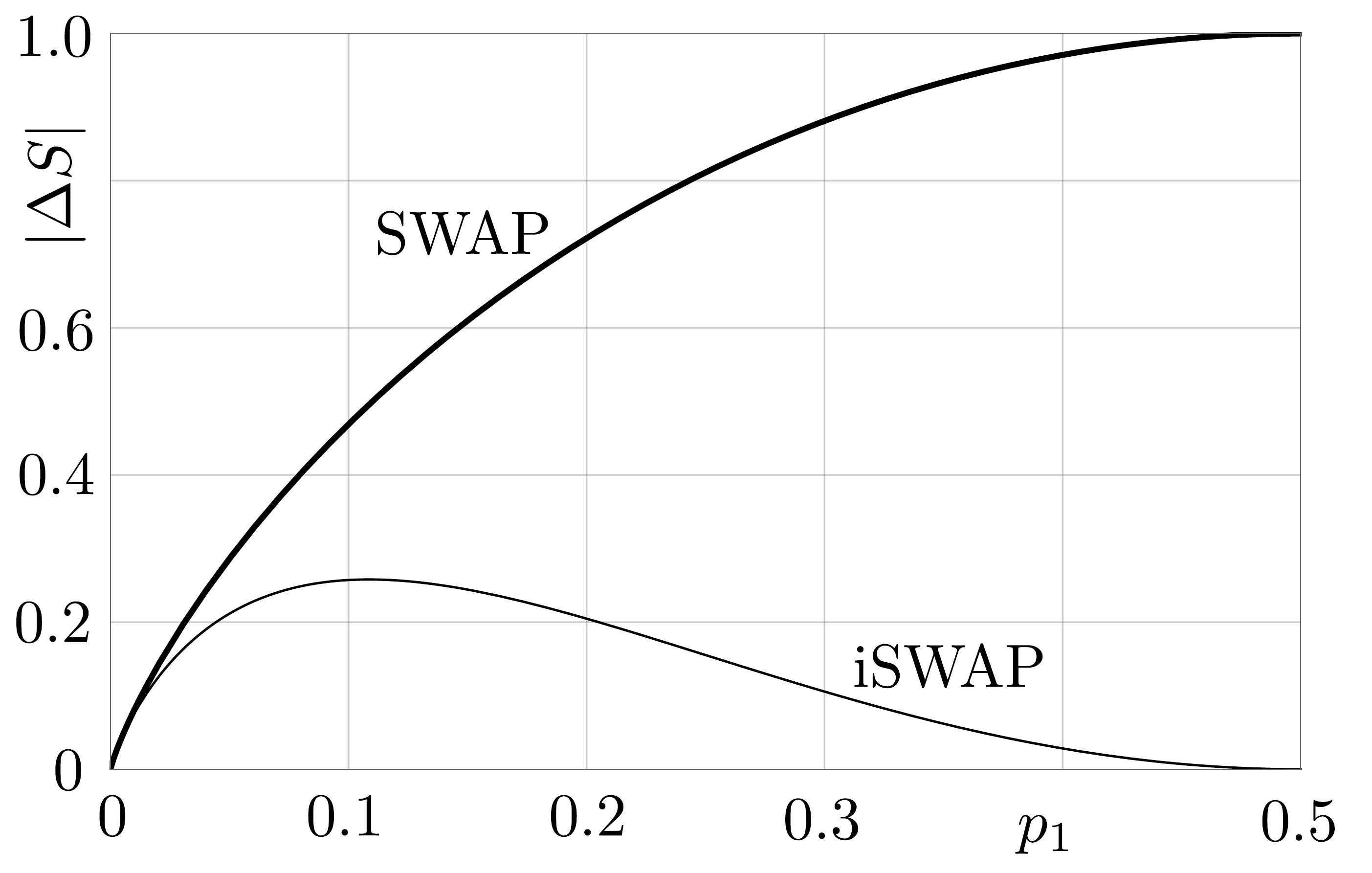}
\caption{Comparison of ideal entropy gain $\Delta S \leq 0$ of a thermal-state
target qubit for the SWAP (thick line) and iSWAP (thin line) operation with a
pure demon qubit as a function of the excited level occupation $p_1$. As $p_1
\to 0$, the target is already pure and the entropy gain vanishes.}
\label{fig:gain}
\end{figure}
\begin{figure*}[t!]
\centering
\includegraphics[width=16cm]{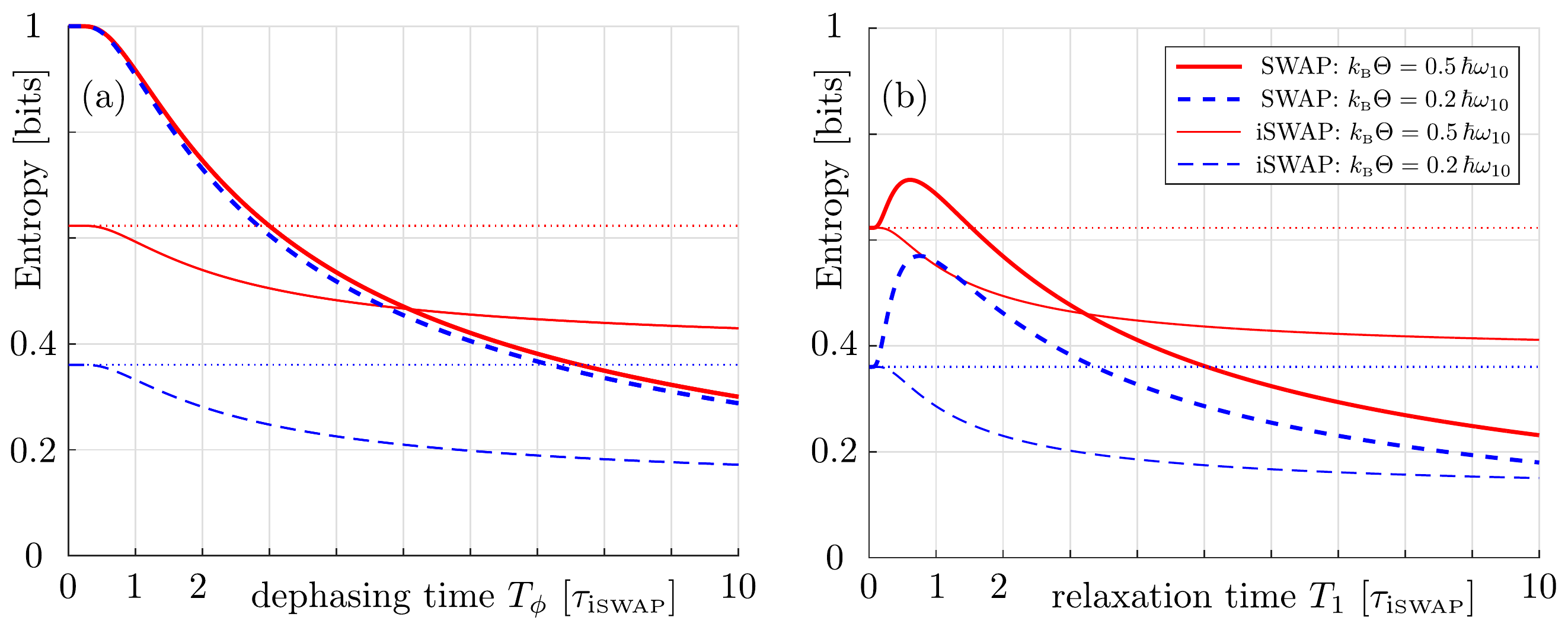}
\caption{(a) Final entropy of the target qubit (vertical axis in bits) after
the execution of the SWAP-QMD (thick lines) and the iSWAP-QMD (thin lines) as
a function of pure dephasing time $T_\phi$ (horizontal axis in units of
$\tau_\mathrm{i{\scriptscriptstyle SWAP}} = h/4J$, relaxation effects are
neglected).  The color/dashes indicate different temperatures of the qubit
environment and the dotted lines mark the level of the target qubit's initial
entropy. (b) Final entropy of the target qubit as a function of relaxation
time $T_1$ in the absence of dephasing effects.} \label{fig:SWAP_vs_iSWAP}
\end{figure*}

However, in practice the qubits are not ideal and prone to decoherence, thus
restricting the available time required for the QMD operation. A way to relax
this time restriction is to replace the SWAP gate by the naturally appearing
iSWAP operation: since a full SWAP involves three $\sqrt{\mathrm{iSWAP}}$
operations, an iSWAP demon performs its task at least 1.5 times faster, which
is quite beneficial given the time constraints due to decoherence.
Furthermore, as shown below, the iSWAP demon is less affected by decoherence.
On the other hand, one has to admit that the iSWAP demon comes with a reduced
purification power\cite{entangling}, see Fig.\ \ref{fig:gain}: starting out
with a thermal state of the target qubit $\hat\rho_\mathrm{th} = p_0 |0\rangle
\langle 0| + p_1 |1\rangle \langle 1|$ and an equal-energy pure state
$\hat\rho_\mathrm{p} = |\chi_\mathrm{th}\rangle \langle \chi_\mathrm{th}|$
with $|\chi_\mathrm{th}\rangle = \sqrt{p_0} |0\rangle + \sqrt{p_1} |1\rangle$
for the demon qubit, the iSWAP-QMD generates a non-entangled but classically
correlated state of the two qubits,
\begin{align}
      & \hat{U}(\tau_\mathrm{i{\scriptscriptstyle SWAP}})\bigl( \hat\rho_\mathrm{th}
        \otimes \hat\rho_\mathrm{p}\bigr)
	\overset{\tau_\mathrm{i{\scriptscriptstyle SWAP}}}{\to}
      \nonumber\\
      & p_0 \,[|\psi_-\rangle \langle \psi_-|] \otimes |0\rangle \langle 0|
	+ p_1\, [|\psi_+\rangle \langle \psi_+| \otimes |1\rangle \langle 1|],
\end{align}
where $|\psi_\pm\rangle = \sqrt{p_0} |0\rangle \pm i \sqrt{p_1}|1\rangle$. The
resulting entropy of the final target state
$\hat\rho(\tau_\mathrm{i{\scriptscriptstyle SWAP}}) = p_0 |\psi_-\rangle
\langle \psi_-| + p_1 |\psi_+\rangle \langle \psi_+|$ is always bounded by the
entropy of its original state; with $S[\hat\rho]$ the von Neumann entropy of
the state $\hat\rho$, the iSWAP QMD provides an entropy reduction $\Delta S =
S[\hat\rho(\tau_\mathrm{i{\scriptscriptstyle SWAP}})] - S[\hat
\rho_\mathrm{th}] \leq 0$ with the equal sign realized for the chaotic state
with $p_0 = p_1 = 1/2$. Note that, during the operation of the iSWAP gate, the
average energy $p_1 \hbar\omega_{1,0}$ of the target qubit remains constant.

As announced above, dephasing and relaxation affect the iSWAP- and SWAP-QMDs
quite differently, with the iSWAP-QMD performing better at short
dephasing/relaxation times, while the SWAP-QMD ultimately outperforms the
iSWAP-QMD at long dephasing/relaxation times due to its higher purification
power. Including dephasing and relaxation in the demon's evolution, we have to
replace the unitary $\hat{U}(\tau)$ in \eqref{eq:unitary} by a channel
$\Phi(\tau)$ that accounts for the environment, see Appendix
\ref{app:Lindblad}. We assume our qubits to interact with their local
environments, each in thermal equilibrium at some temperature $\Theta$, and
adopt the above initial states $\hat{\rho}_\mathrm{th}$ and
$\hat{\rho}_\mathrm{p}$ for the target and demon qubits with $p_1/p_0 =
\exp(-\hbar\omega_{1,0}/k_{\rm \scriptscriptstyle B} \Theta)$. Solving the
corresponding Lindblad master equations \cite{Lindblad} numerically, we
determine the time evolution of the entropies for both demons and for the two
cases of pure dephasing and relaxation.  In Fig.\ \ref{fig:SWAP_vs_iSWAP}, we
show the target qubit's final entropy for different ratios
$T_\phi/\tau_\mathrm{i{\scriptscriptstyle SWAP}}$ (pure dephasing, in (a)) and
$T_1/\tau_\mathrm{i{\scriptscriptstyle SWAP}}$ (relaxation, in (b)) and for
different temperatures.

As expected, short dephasing and relaxation times destroy the purification
power of both demons.  Increasing the dephasing and relaxation times in Fig.\
\ref{fig:SWAP_vs_iSWAP}, we find that the iSWAP-QMD performs better at small
$T_1$ and $T_\phi$, while good qubits with large $T_1$ and $T_\phi$ profit
from the better purification power of the SWAP operation.  An additional
surprise is that the SWAP-QMD fails completely at small $T_1$ and $T_\phi$
where the final entropy turns out higher (and even maximal for
$T_\phi/\tau_\mathrm{i{\scriptscriptstyle SWAP}} \to 0$ in Fig.\
\ref{fig:SWAP_vs_iSWAP}(a)) than the initial one. We attribute this entropy
increase to the action of the intermediate single qubit rotations in the SWAP
operation and its extreme sensitivity to decoherence.

Indeed, consider the extreme case of strong dephasing with $T_\phi$ much
shorter than the duration of the iSWAP gate, $T_\phi \ll
\tau_\mathrm{i{\scriptscriptstyle SWAP}}$, but longer or compatible with the
duration of one-qubit operations. Given such a strong dephasing leads to a
rapid collapse of the qubits to a product of thermal states,
$\hat\rho_\mathrm{th} \otimes \hat\rho_\mathrm{p} \overset{\sqrt{\mathrm{iSWAP}}}{\to}
\hat\rho_\mathrm{th} \otimes \hat\rho_\mathrm{th}$, implying that the entropy
of the target qubit is not changed, see Fig.\ \ref{fig:SWAP_vs_iSWAP}. Going
on with the SWAP-demon, the subsequent $\pi/2$-rotations $\hat{U}_x$ take the
(now thermal) qubits out of the decoherence-free subspace,
$\hat\rho_\mathrm{th} \to \hat{\rho}_\mathrm{rot} = \hat{U}_x
\hat\rho_\mathrm{th} \hat{U}_x^\dagger \equiv (1/2)\mathds{1} + \hat\sigma_y
(p_0-p_1)/2$, and the subsequent evolution brings both qubits into the
maximally mixed state, $\hat{\rho}_\mathrm{rot} \otimes
\hat{\rho}_\mathrm{rot} \overset{\sqrt{\mathrm{i{\scriptscriptstyle
SWAP}}}}{\to} (1/2)\mathds{1} \otimes (1/2) \mathds{1}$.

The same argument is valid for strong relaxation $T_1 \to 0$: the evolution
rapidly takes the pair of qubits into a product of thermal states, while
subsequent $\hat{U}_x$ rotations take them out of the equilibrium with their
local environment, $\hat\rho_\mathrm{th} \to \hat{\rho}_\mathrm{rot}$. During
the next $\sqrt{\mathrm{iSWAP}}$ gate, both qubits relax back to the thermal
equilibrium state, $\hat{\rho}_\mathrm{rot} \otimes \hat{\rho}_\mathrm{rot}
\overset{\sqrt{\mathrm{i{\scriptscriptstyle SWAP}}}}{\to} \hat\rho_\mathrm{th}
\otimes \hat\rho_\mathrm{th}$, keeping the target qubit in the original
entropy state.  However, for a moderately strong relaxation strength, the
state $\hat{\rho}_\mathrm{rot}$ may not have enough time to relax back into
the thermal state $\hat\rho_\mathrm{th}$ during the square-root iSWAP time
$\tau_\mathrm{i{\scriptscriptstyle SWAP}}/2$ and thus ends up in a higher
entropy state. This explains the non-monotonic behavior of the entropy of the
target qubit at short relaxation times in Fig.\ \ref{fig:SWAP_vs_iSWAP}(b).
The above discussion lets us conclude that the iSWAP-QMD is less sensitive to
dephasing and relaxation than its SWAP analog; only for long dephasing and
relaxation times $T_\phi, T_1 \gg \tau_\mathrm{i{\scriptscriptstyle SWAP}}$
does the better purification power of the SWAP gate beat the preformance of
the iSWAP-demon.

\section{Operational requirements for the demon}\label{sec:Dec_Pur}

The functionality of the iSWAP demon depends critically on its environment, of
which the transmission line is an integral part. A large separation between
the target and demon qubits reduces the coupling $J$ and thus enhances the
operation time $\tau_\mathrm{i{\scriptscriptstyle SWAP}}$, what is in conflict
with the finite decoherence time $T_2$; as a result, we obtain a limit $\ell$
of the demon's extension, see Sec.\ \ref{sec:demon_ex}. Second, the presence
of the transmission line will itself increase both the dephasing and the
relaxation rate of the qubits.  While dephasing due to the thermally excited
bosonic modes will limit the operational temperature of the transmission line,
see Sec.\ \ref{sec:Deph}, we find that the enhanced qubit relaxation due to a
lossy transmission line (Purcell effect, see Sec.\ \ref{sec:Rel}) is not (yet)
relevant in our setup.

\subsection{Demon extension}\label{sec:demon_ex}

Given a decoherence time $T_2 = (1/2T_1 + 1/T_\phi)^{-1}$, we estimate the
possible extension $\ell$ of the demon.  In order to successfully realize an
iSWAP operation, the condition $\tau_\mathrm{i{\scriptscriptstyle SWAP}} \sim T_2$ has to be
satisfied.  Choosing the specific arrangement $\omega_{1,0} = 2\omega_0$ for
the resonator and qubit transition frequencies and similar target and demon
qubits with $\kappa^{\rm \scriptscriptstyle L} = \kappa^{\rm
\scriptscriptstyle R}= \kappa$, one finds that
\begin{equation}
       T_2\sim \frac{\ell}{v} \, \frac{1}{8\kappa^2}
       \frac{\omega_0}{\Delta\omega}.
\end{equation}
Assuming typical values $v \sim 2c/3$, $\omega_0 \sim 2\pi \times 5$ GHz, and
$\Delta\omega \sim 2\pi\times 500$ MHz, we find that typical coherence times
$T_2 \sim 50 - 250~\mu$s allow for an extension of the demon over macroscopic
lengths $\ell \sim 1.0 - 5.0$ meters.

\subsection{Dephasing and relaxation due to transmission line}

\subsubsection{Dephasing and transmission line temperature}\label{sec:Deph}

So far, we have assumed that the transmission line is kept at low temperature
such that electromagnetic modes within its bandwidth $[\omega_0 -
{\Delta\omega}/2, \omega_0 + {\Delta\omega}/2]$ are not thermally excited.
Given the possibility for a macroscopic separation $\ell$, a question of much
technological interest then is, whether the QMD can be operated through a hot
and thus less-quantum environment. Indeed, at finite temperatures a thermal
voltage noise appears in the transmission line that causes dephasing of the
qubits. The dephasing due to the presence of a (hot) transmission line is
described via the dispersive shift of the qubit energy levels induced by the
fluctuating voltages at the ends of the transmission line.  The corresponding
qubit-dephasing Hamiltonian is given by $\hat{H}_\mathrm{sh} =
\sum_{\alpha={\rm \scriptscriptstyle L,R}} \bigl[|0\rangle_{\!\alpha}\,
{}_{\alpha\!} \langle 0| - |1\rangle_{\!\alpha}\, {}_{\alpha\!} \langle
1|\bigr] \otimes \hat{B}^\alpha(x^\alpha)$, where the operators
$\hat{B}^\alpha(x^\alpha) = \sum_{n,m} b_{nm}(x^\alpha) \hat{a}_n^\dagger
\hat{a}_m$ describe the coupling to the transmission line modes of the
electromagnetic environment, see Appendix \ref{app:Int_Ham} (here, $\hat{a}_n$
denote bosonic operators of the transmission line). The presence of thermal
modes in the transmission line then modifies the level separation of the qubit
and induces dephasing at a rate (see Appendix \ref{app:Dephas})
\begin{eqnarray}
      &&\gamma_\phi^{\alpha\alpha^\prime}
      =32\pi\, (\kappa^\alpha \kappa^{\alpha^\prime})^2
	N_{\omega_0}(1\!+\!N_{\omega_0}) \,\Delta\omega
      \\
      &&\times\biggl[
      \frac{\omega_\mathrm{an}^{\alpha}\omega_0}
	{(\omega_{1,0}^{\alpha}-\omega_0)
	(\omega_{2,1}^{\alpha}-\omega_0)}\biggr]
      \biggl[ \frac{\omega_\mathrm{an}^{\alpha^\prime }\omega_0}
	{(\omega_{1,0}^{\alpha^\prime}-\omega_0)
	(\omega_{2,1}^{\alpha^\prime}-\omega_0)}\biggr],
      \nonumber
\end{eqnarray}
where $N_\omega$ is the Bose-Einstein distribution function and
$\omega_\mathrm{an}^{\alpha} = \omega_{1,0}^{\alpha} - \omega_{2,1}^{ \alpha}$
are the anharmonicities of the transmon spectra. Remarkably, the dephasing rate
scales as $\gamma_\phi \propto \kappa^4$, while the coupling constant $J
\propto \kappa^2$, see Eq.\ (\ref{eq:coupling}); both of them are linear in
the frequency bandwidth $\Delta\omega$. Therefore, installing a small coupling
$\kappa$ one can keep the qubit dephasing rates small while leaving a
sufficiently strong coupling between the qubits. The iSWAP-QMD device is
functional when $\gamma_\phi\tau_\mathrm{i{\scriptscriptstyle SWAP}} \leq 1$, that translates into
a requirement on the photon occupation number
\begin{equation}
      N_{\omega_0}(1+N_{\omega_0}) \leq \biggl[ 8\pi \kappa^2 \frac{\omega_0 \ell}{v}
	\frac{\omega_0}{\omega_{1,0}} \Bigl(\frac{\omega_\mathrm{an}}
	{\omega_{2,1}-\omega_0}\Bigr)^2\biggr]^{-1}.
\end{equation}
For $\omega_0 \sim 2\pi \times 5$ GHz, $v\sim 2c/3$, and $\omega_\mathrm{an} \sim
2\pi \times 300$ MHz, we find that $N_{\omega_0} (1+N_{\omega_0}) \leq
1000~\mathrm{m}/\ell$, that translates to a corresponding temperature range
$7.5~\mathrm{K} \geq \Theta_\mathrm{line} \geq 3.5~\mathrm{K}$ for $1~\mathrm{m}
\leq \ell \leq 5.0~\mathrm{m}$. Hence, the transmission line can reside at a
temperature that is about two orders of magnitude higher than the typical
operation temperature $\Theta_\mathrm{qubit} \sim 20$ mK of the superconducting
qubits.

\subsubsection{Relaxation through Purcell effect}\label{sec:Rel}

Finally, we study the consequences of losses in the transmission line. Indeed,
a finite loss rate $\gamma_\mathrm{line}$ in the transmission line induces an
enhanced decay of the qubit excited state via the Purcell effect (see Appendix
\ref{app:Rel}),
\begin{equation}
      \gamma_{\rm Pur}^{\alpha}
      = 2\, \gamma_\mathrm{line} \, (\kappa^\alpha)^2 \frac{\Delta\omega\, \omega_0}
	{(\omega_{1,0}^{\alpha}\!-\!\omega_0)^2}.
\end{equation}
For commercially available coaxial cables with an attenuation constant $\sim
0.1$ dB/m one has $\gamma_\mathrm{line} \sim 4.6$ MHz and choosing parameters
as above results in a lifetime $\gamma_{\rm Pur}^{-1} \sim 9$ ms that is long
compared to the relaxation time $T_1$ assumed above. Hence, we conclude that
the presence of the transmission line does not significantly reduce the
performance of the qubit's characteristics that we have assumed above.

\section{Conclusion}\label{sec:conclusion}

In conclusion, we have proposed a realistic design for a spatially distributed
quantum Maxwell demon based on a cQED platform.  The transmon-type target and
demon qubits are capacitively coupled via the electromagnetic modes of a
transmission line; its non-resonant coupling allows to keep the line at high
temperatures, of order Kelvin, while the resonant coupling of other designs
\cite{Wallraff:2018} requires a cold line.  The device serves to reduce the
entropy of the target qubit via exchange of its state with a higher-purity
demon state. Previous demons, both local \cite{Lloyd:1997} and extended
\cite{lebedev:2016}, were based on SWAP or partial-SWAP operations involving
multiple CNOT gates; here, we have proposed to reduce the demon's complexity
via operating on the `machine code' level by directly exploiting the XY-type
coupling between the qubits.  The resulting iSWAP gate then provides limited
purification power to the demon but behaves more benevolent with respect to
decoherence.  Our estimates show, that the target qubit can be purified `from
a distance', with the demon qubit located a macroscopic distance of order
meters away.  The proposed setup can be implemented with present day
technology.

\acknowledgements

This work was supported by the Swiss National Foundation through the NCCR QSIT (A.V.L.), the U.S.\ Department of Energy, Office of Science, Basic Energy Sciences, Materials Sciences and Engineering Division (A.V.L., V.M.V.),  the RFBR Grant 18-02-00642A, the Foundation for the Advancement of Theoretical Physics BASIS (G.B.L.), the Ministry of Education and Science of the Russian Federation 16.7162.2017/8.9 (A.V.L.), and the Government of the Russian Federation through the Agreement 05.Y09.21.0018 (G.B.L., A.V.L.).

\appendix

\section{Transmission line}\label{app:Trans_Line}

An ideal lossless transmission line can be modelled as a pair of uniform
conductors separated by a dielectric medium; it is characterized by a series
of inductances ${\cal L}$ (in Henry/meter) and shunt capacitances ${\cal C}$
(Farad/meter) \cite{pozar:2012}. The voltage $V(x,t)$ and current $I(x,t)$
along the transmission line is described by the transmission line equations
\begin{equation}
   \partial_x V = - {\cal L}\,\partial_t I,
   \qquad
   \partial_x I = - {\cal C}\,\partial_t V.
\end{equation}
Introducing the potential $\varphi(x,t)$ and expressing the voltage and
current via $V(x,t) = \sqrt{\cal L}\, \partial_t \varphi(x,t)$ and $I(x,t) =
-\partial_x \varphi(x,t)/\sqrt{\cal L}$, the transmission line equations
reduce to a standard wave equation (with $\dot\varphi = \partial_t \varphi$
and $\varphi' = \partial_x \varphi$),
\begin{equation}
      \ddot{\varphi} - c^2 \varphi^{\prime\prime} = 0,
      \label{eq:wave}
\end{equation}
where $c=1/\sqrt{\cal LC}$ is the wave velocity.

The quantization of the transmission line fields \cite{She:1965} is done via
standard canonical quantization. The classical equation of motion
(\ref{eq:wave}) derives from minimizing the classical action $S =
\int dt\, dx\, L(\dot\varphi,\varphi)$ with the Lagrangian density
\begin{equation}
    L(\dot\varphi,\varphi) = \frac12 \bigl[ \bigl( \dot\varphi/c
    \bigr)^2 - \bigl( \varphi^\prime \bigr)^2 \bigr].
\end{equation}
Introducing the conjugated field $\pi(x) = \partial L/\partial \dot\varphi =
\dot\varphi/c^2$, provides us with the Hamiltonian $H(\pi,\varphi) =
\int dx\, \pi(x) \dot\varphi(x) - L(\pi,\varphi)$,
\begin{equation}
    H(\pi,\varphi) = \frac12 \int dx \bigl[ c^2\pi^2(x)
    + {\varphi^\prime}^2(x) \bigr].
    \label{eq:lineH}
\end{equation}
Going back to the original fields $V(x) = c^2 \sqrt{\cal L}\, \pi(x)$ and
$I(x) = - \varphi^\prime(x)/\sqrt{\cal L}$, we obtain the transmission line
Hamiltonian
\begin{equation}
      H = \frac12 \int dx \bigl[ {\cal C} V^2(x) + {\cal L} I^2(x) \bigr].
\end{equation}

The canonical quantization maps the classical fields to operators, $\varphi(x)
\to \hat\varphi(x)$ and $\pi(x) \to \hat\pi(x)$ with commutation relations
$[\hat\varphi(x), \hat\varphi(y)] = [\hat\pi(x),\hat\pi(y)] = 0$ and $[
\hat\varphi(x), \hat\pi(y)] = i\hbar\delta(x-y)$. Introducing the transmission
line modes
\begin{eqnarray}
      &&\hat\varphi(x)= c\sum_k \Bigl( \frac{\hbar}{2\omega_k \ell} \Bigr)^{1/2}
        \!\bigl( \hat{a}_k e^{ikx} + \hat{a}_k^\dagger e^{-ikx} \bigr),
      \\
      &&\hat\pi(x)= -\frac{i}{c} \sum_k \Bigl( \frac{\hbar\omega_k}{2\ell} \Bigr)^{1/2}
        \!\bigl( \hat{a}_k e^{ikx} - \hat{a}_k^\dagger e^{-ikx} \bigr),
\end{eqnarray}
we go over to bosonic annihilation and creation operators $\hat{a}_k$ and
$\hat{a}_k^\dagger$ with commutators $[\hat{a}_k, \hat{a}_{k^\prime}^\dagger]
= \delta_{kk^\prime}$, dispersion $\omega_k = c|k|$, and $\ell \to \infty$ is
the length of the transmission line. The Hamiltonian (\ref{eq:lineH}) then
transforms into the standard form
\begin{equation}
      H(\pi,\varphi) \to \hat{H}_\mathrm{line} = \frac12 \sum_k \hbar\omega_k
        \bigl( \hat{a}_k^\dagger \hat{a}_k + \hat{a}_k \hat{a}_k^\dagger \bigr).
\end{equation}
The voltage and current operators derive from the mode operators $\hat{a}_k$
and $\hat{a}_k^\dagger$ via
\begin{eqnarray}
      &&\hat{V}(x) = -i \sum_k \Bigl( \frac{\hbar\omega_k}{2 C_r} \Bigr)^{1/2}
        \bigl( \hat{a}_k e^{ikx} - \hat{a}_k^\dagger e^{-ikx} \bigr)
      \nonumber\\
      &&\qquad \equiv \sum_k \hat{V}_k e^{ikx} + h.c.,
      \\
      &&\hat{I}(x) = -i \sum_k \mbox{sgn}(k) \Bigl( \frac{\hbar\omega_k}{2 L_r}
        \Bigr)^{1/2} \bigl( \hat{a}_k e^{ikx} - \hat{a}_k^\dagger e^{-ikx} \bigr)
      \nonumber\\
      &&\qquad \equiv \sum_k \hat{I}_k e^{ikx} + h.c.,
\end{eqnarray}
where $C_r = {\cal C}\ell$ and $L_r = {\cal L}\ell$ are the total capacitance
and inductance of the transmission line. The $k$-components of the voltage- and
current-operators are linearly related through a transmission line impedance
$Z_0 = \sqrt{\cal L/C}$,
\begin{equation}
      \hat{V}_k = Z_0\, \mbox{sgn}(k) \hat{I}_k.
\end{equation}

For an open transmission line, we have to impose the boundary conditions
$\hat{I}(x=\pm \ell/2) =0$, resulting in a discrete level spectrum with wave
numbers $k_n = \pi n/\ell$, $n\geq 0$, describing even and odd modes
\begin{equation}
      \hat{V}(x) = -i \sum_n \Bigl( \frac{\hbar\omega_n}{C_r} \Bigr)^{1/2}
        \varphi_n(x) \hat{a}_n + h.c.,
      \label{eq:V}
\end{equation}
where
\begin{equation}
      \varphi_n(x) = \left\{ \begin{array}{rl}
      \cos(\pi n x/\ell),&\quad n~ \mathrm{even},\\
      i\sin(\pi n x/\ell),&\quad n~ \mathrm{odd}.
      \end{array}\right.
\end{equation}

\section{Interaction Hamiltonian for transmon qubits}\label{app:Int_Ham}

We wish to eliminate the transmission line modes in the Hamiltonian
\eqref{eq:ham} to lowest order in the bosonic operators $a_n$; this will
provide us with the effective qubit--qubit coupling and higher-order terms
including a dispersive shift describing transmission-line induced dephasing.
We use a perturbative scheme that is valid in the off-resonant regime
$|\omega_{1,0}^{ \alpha} - \omega_0|\sim \omega_0$. We perform a unitary
transformation, $\hat{H} \to \hat{\cal H} = \hat{U}\hat{H} \hat{U}^\dagger$
with $\hat{U} = \exp\bigl[ \hat{S} - \hat{S}^\dagger \bigr]$, and seek an
operator $\hat{S} = \sum_{\alpha,i} q_{i+1,i}^{\alpha}
|i+1\rangle_{\!\alpha}\, {}_{\alpha\!}\langle i|\, \hat{Q}^\alpha_i$ that
eliminates the terms linear in $a_n$ within the expansion $\hat{\cal H}
\approx \hat{H} + [\hat{S},\hat{H}] + [\hat{S},\hat{H}]^\dagger +\dots$. This
is achieved by the choice
\begin{equation}
   \hat{Q}^\alpha_i=-i\sum_n \Bigl( \frac{\omega_n}{\hbar C_r} \Bigr)^{1/2}
   \frac{\varphi_n(x_\alpha)}{\omega_{i+1,i}^{\alpha}
   - \omega_n}\, \hat{a}_n.
      \label{eq:Q}
\end{equation}
The transformed Hamiltonian then takes the form
\begin{equation}
   \hat{\cal{H}} \approx \hat{H}_\mathrm{transmon} + \hat{H}_\mathrm{bath}
   + \hat{H}_\mathrm{sh} + \hat{H}_\mathrm{2ph} + \hat{H}_\mathrm{int},
\end{equation}
where $\hat{H}_\mathrm{bath} = \sum_n \hbar\omega_n \hat{a}_n^\dagger
\hat{a}_n$ is the transmission line Hamiltonian and $\hat{H}_\mathrm{sh}$
describes the dispersive shift of the transmon's energy levels due to
the off-resonant interaction with the transmission line modes,
\begin{eqnarray}
      &&\hat{H}_\mathrm{sh} = |q_{i+1,i}^{\alpha}|^2
        \sum_{i,\alpha} |i\!+\!1\rangle_{\!\alpha}\,{}_{\alpha\!}\langle i\!+\!1|
        \bigl( \hat{Q}^\alpha_i [\hat{V}^\alpha]^\dagger + \hat{V}^\alpha
        [\hat{Q}^\alpha_i]^\dagger \bigr)
      \nonumber
      \\
      &&\qquad - |q_{i+1,i}^{\alpha}|^2
        \sum_{i,\alpha} |i\rangle_{\!\alpha}\,{}_{\alpha\!}\langle i|
        \bigl([\hat{V}^\alpha]^\dagger
        \hat{Q}^\alpha_i + [\hat{Q}^\alpha_i]^\dagger \hat{V}^\alpha \bigr).
      \label{eq:Hshift}
\end{eqnarray}
The contribution $\hat{H}_\mathrm{2ph}$ describes the next-order two-photon
interaction process,
\begin{equation}
      \hat{H}_\mathrm{2ph} =  \sum_{\alpha,i} |i\!+\!2\rangle_{\!\alpha}\,{}_{\alpha\!}
       \langle i| \otimes \hat\eta_i^{\alpha} + h.c.,
\end{equation}
where $\hat\eta_i^{\alpha} = q_{i+i,i}^{
\alpha} \, q_{i+2,i+1}^{\alpha} \bigl( \hat{Q}^\alpha_{i+1} -
\hat{Q}^\alpha_i\bigr)\hat{V}^\alpha$. Finally, the term
$\hat{H}_\mathrm{int}$ describes the directly induced interaction between the
transmon qubits,
\begin{eqnarray}
    &&\hat{H}_\mathrm{int} = \sum_{\alpha\neq \beta} \sum_{i,j}
    |i\!+\!1\rangle_{\!\alpha}\,{}_{\alpha\!} \langle i|
    \otimes |j\rangle_{\!\beta}\,{}_{\beta} \langle j\!+\!1|
    \label{eq:Hint_comm}\\
    &&\qquad\quad \times q_{i+1,i}^{\alpha}
    q_{j,j+1}^{\beta} \, \bigl[ \hat{Q}^\alpha_i,
    [\hat{V}^\beta]^\dagger\bigr] + h.c.
      \nonumber
\end{eqnarray}

The effective coupling constant involves the commutator of the
electromagnetic field operators at the opposite ends of the transmission
line. Making use of the explicit form of the operators $\hat{V}^\alpha$ and
$\hat{Q}^\alpha_i$, see Eqs.\ (\ref{eq:V}) and (\ref{eq:Q}), one finds,
\begin{eqnarray}
      &&\bigl[ \hat{Q}^\alpha_i, [\hat{V}^\beta]^\dagger \bigr]
        = \sum_{n~\mathrm{even}} \frac{\omega_n}{C_r}
      \frac{\cos(\pi n x^\alpha/\ell) \cos(\pi n x^\beta/\ell)}
      {\omega_{i+1,i}^{\alpha} - \omega_n}
      \nonumber
      \\
      &&\qquad + \sum_{n~\mathrm{odd}} \frac{\omega_n}{C_r}
     \frac{\sin(\pi n x^\alpha/\ell) \sin(\pi n x^\beta/\ell)}
     {\omega_{i+1,i}^{\alpha} - \omega_n}.
\end{eqnarray}
In particular, for transmon qubits located at the opposite ends
$x^\alpha = -x^\beta = \ell/2$ of the transmission line, one has
\begin{equation}
      \bigl[ \hat{Q}^\alpha_i, [\hat{V}^\beta]^\dagger \bigr]
        = \frac1{C_r} \sum_{k=1}^\infty \frac{(\omega_{2k}-\omega_{2k-1})
      \, \omega_{i+1,i}^{\alpha}}{
      (\omega_{i+1,i}^{\alpha}-\omega_{2k})
      (\omega_{i+1,i}^{\alpha} - \omega_{2k-1})}.
\end{equation}
Going to the continuous limit $\sum_k \to \int \frac{\ell d\omega}{2\pi c}$
one finds,
\begin{equation}
    \bigl[\hat{Q}^\alpha_i, [\hat{V}^\beta]^\dagger \bigr]
    = \frac1{2C_r} \int d\omega \, \frac{\omega_{i+1,i}^{\alpha}}
    {(\omega_{i+1,i}^{\alpha}-\omega)^2}.
\end{equation}
Substituting this expression into Eq.\ (\ref{eq:Hint_comm}), one finally
arrives at the qubit--qubit coupling constants $J_{ij}$ given in Eq.\
\eqref{eq:Jij} of the main text.

\section{Phenomenological Lindblad Analysis}\label{app:Lindblad}

\subsection{Qubit relaxation} \label{app:Qubit_Rel}

We assume that each qubit interacts with its local environment and describe
the evolution of the two-qubit density matrix $\hat\rho(t)$ by the Lindblad
equation,
\begin{equation}
      \frac{d\hat\rho(t)}{dt} = -i\bigl[ \hat{H}(t), \hat\rho(t) \bigr]
      + \sum_{\alpha = {\rm \scriptscriptstyle L,R}}
      {\cal D}_\mathrm{rel}^\alpha[\hat\rho(t)],
      \label{eq:Lindblad}
\end{equation}
where $\hat{H}(t)$ is the time-dependent Hamiltonian which describes the
sequence of a one- and two-qubit operations applied during the execution of
the SWAP and iSWAP Maxwell demon and ${\cal D}_\mathrm{rel}^\alpha[\hat\rho]$
is the dissipator that describes the excitation/relaxation processes for the
qubits $\alpha = \mathrm{L,R}$,
\begin{equation}
      {\cal D}_\mathrm{rel}^\alpha[\hat\rho] = \sum_{\mu = \pm} \gamma_\mu^\alpha
        \Bigl( \hat{\sigma}_\mu^\alpha\, \hat\rho\, [\hat{\sigma}_\mu^\alpha]^\dagger
        - \frac12 \bigl\{ [\hat{\sigma}_\mu^\alpha]^\dagger \hat{\sigma}_\mu^\alpha,
        \hat\rho\bigl\} \Bigr).
      \label{eq:Dissipator}
\end{equation}
Here, $\hat{\sigma}_-^\alpha = |0\rangle_{\!\alpha} \, {}_{\alpha\!} \langle
1|$ and $\hat{\sigma}_+^\alpha = |1\rangle_{\!\alpha} \, {}_{\alpha\!}\langle
0|$ describe the relaxation and excitation processes with rates
$\gamma_\pm^\alpha > 0$ and  $\{\cdot,\cdot\}$ is the anti-commutator.  If
both local environments are in thermal equilibrium at a temperature $\Theta$,
then $\gamma_+^\alpha = \gamma_-^\alpha \exp(-\beta\hbar\omega_{1,0})$ with
$\beta = 1/k_{\rm \scriptscriptstyle B}\Theta$.

We assume that the target qubit on the left initially is in thermal
equilibrium with its environment, $\hat\rho_\mathrm{t}(0) = \hat\rho_\mathrm{th}
= p_0 |0\rangle_{ \rm \scriptscriptstyle L}\, {}_{\rm\scriptscriptstyle
L\!}\langle 0| + p_1 |1\rangle_{\rm \scriptscriptstyle L}\,
{}_{\rm\scriptscriptstyle L\!} \langle 1|$, where $p_0 = [1+
\exp(-\beta\hbar\omega_{1,0})]^{-1}$ and $p_1 = [1+
\exp(\beta\hbar\omega_{1,0})]^{-1}$ are equilibrium occupation probabilities.
In contrast, the demon qubit on the right is prepared in the equal-energy pure
state, $\hat\rho_\mathrm{d}(0) = \hat\rho_\mathrm{p} = |\chi_0\rangle \langle
\chi_0|$ with $|\chi_0\rangle = \sqrt{p_0}\, |0\rangle_{\rm \scriptscriptstyle
R} + \sqrt{p_1}\, |1\rangle_{\rm \scriptscriptstyle R}$.

The execution of the iSWAP operation in the presence of the relaxation
processes is described by Eq.\ (\ref{eq:Lindblad}) with the constant
Hamiltonian,
\begin{equation}
      \hat{H}(t) = J\left[ \begin{array}{cccc}
      1&0&0&0\\
      0&0&1&0\\
      0&1&0&0\\
      0&0&0&1
      \end{array} \right]
      \label{eq:HiSWAP}
\end{equation}
acting during the time interval $0<t\leq \tau_\mathrm{i{\scriptscriptstyle SWAP}} = h/4J$. The
Lindblad equation (\ref{eq:Lindblad}) with the Hamiltonian (\ref{eq:HiSWAP})
describes the linear evolution of the $16$ components of the density matrix
$\hat\rho(t)$ and its formal result can be written in the form of a quantum
channel,
\begin{equation}
   \hat\rho(t=\tau_\mathrm{i{\scriptscriptstyle SWAP}}) = \Phi_\mathrm{iSWAP}\bigl[ \hat\rho_\mathrm{t}(0)
   \otimes \hat\rho_\mathrm{d}(0) \bigr].
\end{equation}
For its numerical solution, we assume that $\gamma_{\pm,{\rm
\scriptscriptstyle L}} = \gamma_{\pm,{\rm \scriptscriptstyle R}} =
\gamma_\pm$, such that the result merely depends on the two dimensionless
parameters $\gamma_-/J$ and $\beta \hbar\omega_{1,0}$.

On the other hand, the SWAP demon involves the consecutive
transformations of the density matrix $\hat\rho$ according to the quantum
circuit shown in Fig.\ 2 of the main text,
\begin{eqnarray}
   \hat\rho(0) &\to& \hat\rho_1^\prime =\Phi_{\sqrt{i\mathrm{SWAP}}}|\bigl[ \hat\rho(0) \bigr]
   \\
   &\to& \hat\rho_1 = \bigl[\hat{U}_x\otimes \hat{U}_x\bigr]
   \cdot \hat\rho_1^\prime \cdot \bigl[\hat{U}_x^\dagger \otimes \hat{U}_x^\dagger\bigr]
   \nonumber\\
   &\to& \hat\rho_2^\prime = \Phi_{\sqrt{i\mathrm{SWAP}}}\bigl[ \hat\rho_1 \bigr]
   \nonumber\\
   &\to& \hat\rho_2 = \bigl[\hat{U}_y\hat{U}_x^\dagger \otimes \hat{U}_y\hat{U}_x^\dagger\bigr]
   \cdot \hat\rho_2^\prime \cdot \bigl[\hat{U}_x \hat{U}_y^\dagger \otimes \hat{U}_x \hat{U}_y^\dagger\bigr]
   \nonumber\\
   &\to& \hat\rho_3^\prime = \Phi_{\sqrt{i\mathrm{SWAP}}}\bigl[ \hat\rho_2 \bigr]
   \nonumber\\
   &\to& \hat\rho_3 = \bigl[\hat{U}_y^\dagger\otimes \hat{U}_y^\dagger\bigr]
   \cdot \hat\rho_3^\prime \cdot \bigl[\hat{U}_y \otimes \hat{U}_y\bigr],
      \nonumber
\end{eqnarray}
where $\hat{U}_x = \exp[-i\pi\, \hat\sigma_x/4]$ is a spin-$1/2$ rotation by
$\pi/2$ around $x$-axis, $ \Phi_{\sqrt{i\mathrm{SWAP}}}$ is a quantum channel, corresponding to $\sqrt{\mathrm{iSWAP}}$ execution in the presence of decoherence. In the above transformation, we have assumed that the
one-qubit rotations take a negligible time in comparison with the $\sqrt{\mathrm{iSWAP}}$
operation and therefore the relaxation processes can be neglected during their
execution. In Fig.\ \ref{fig:SWAP_vs_iSWAP}(b), we show the von Neumann entropy $S[\hat\rho]$
evaluated for the SWAP and iSWAP channels for a qubit evolution including
relaxation processes characterized by $\gamma_- = 1/T_1$.

\subsection{Qubit dephasing}\label{app:Qubit_Deph}

Dephasing is phenomenologically accounted for by the dissipator
\begin{equation}
   {\cal D}_\mathrm{dph}[\hat\rho]
   = \sum_{\alpha,\alpha^\prime} \gamma_\phi^{\alpha
   \alpha^\prime} \bigl( \hat\sigma_\mathrm{z}^{\alpha}
   \hat\rho\, \hat\sigma_\mathrm{z}^{\alpha^\prime}
   - \bigl\{ \hat\sigma_\mathrm{z}^{\alpha^\prime}
   \hat\sigma_\mathrm{z}^{\alpha},\hat\rho\bigr\}/2\bigr),
\end{equation}
where $\hat\sigma_\mathrm{z}^{\alpha} = |0\rangle_{\!\alpha} \,{}_{\alpha\!}
\langle 0| - |1\rangle_{\!\alpha} \,{}_{\alpha\!} \langle 1|$ and
$\gamma_\phi^{ \alpha\alpha^\prime}$ are pure dephasing rates. In our
numerical evaluation of the channel $\Phi(t)$, we use the phenomenological
parameter $\gamma_\phi^{\alpha \alpha^\prime} = 1/T_\phi$.  In Fig.\
\ref{fig:SWAP_vs_iSWAP}(a), we show the von Neumann entropy $S[\hat\rho]$
evaluated for the SWAP and iSWAP channels for a qubit evolution with pure
dephasing processes.

\section{Transmission-line induced decoherence}\label{app:TL_Decoh}

The transmission line induces qubit dephasing due to the presence of thermal
photons and we present a microscopic analysis of this effect in Appendix
\ref{app:Dephas}. Furthermore, a lossy transmission line enhances the qubit's
relaxation rate via the Purcell effect which is studied in Appendix \ref{app:Rel}.

\subsection{Qubit dephasing} \label{app:Dephas}

We determine the transmission-line induced dephasing rate of the qubit. The
latter can be derived microscopically by accounting for the sensitivity of the
qubit's energy levels to the presence of photons in the transmission line, as
described by the dispersive shift $\hat{H}_\mathrm{sh}$ in Eq.\
(\ref{eq:Hshift}). In the qubit subspace, this Hamiltonian can be written as
\begin{equation}
   \hat{H}_\mathrm{sh} = \sum_\alpha \hat\sigma_\mathrm{z}^{\alpha}
   \otimes \hat{B}^\alpha(x^\alpha),
   \label{eq:Hdeph1}
\end{equation}
with the bosonic fields $\hat{B}^\alpha(x^\alpha)$ given by
\begin{eqnarray}
      &&\hat{B}^\alpha(x^\alpha) \! = \! \frac12\sum_{n,m} \Bigl( \frac{\omega_n \omega_m}
        {C_r^2} \Bigr)^{1/2} \! \biggl\{ \frac{|q_{2,1}^{\alpha}|^2}
        {\omega_{2,1}^{\alpha}-\omega_n}
        \! - \! \frac{2|q_{1,0}^{\alpha}|^2}
        {\omega_{1,0}^{\alpha}-\omega_n}
      \nonumber\\
      &&+\frac{|q_{2,1}^{\alpha}|^2}
        {\omega_{2,1}^{\alpha}\!-\omega_m}
        \! - \! \frac{2|q_{1,0}^{\alpha}|^2}
        {\omega_{1,0}^{\alpha}\!-\omega_m}\biggr\}
        \varphi_n^*(x^\alpha) \varphi_m(x^\alpha)\, \hat{a}_n^\dagger \hat{a}_m.
\end{eqnarray}
The second qubit level $i=2$ appears through the terms $i=1$ in the second
line of Eq.\ (\ref{eq:Hshift}).  Using the relation $|q_{2,1}^{\alpha}|^2 = 2
|q_{1,0}^{\alpha}|^2$, this expression simplifies to
\begin{eqnarray}
      &&\hat{B}^\alpha(x^\alpha) \! = \! (\kappa^\alpha)^2 \frac{hv}{\ell}
        \sum_{n,m} \biggl\{\! \frac{\omega_\mathrm{an}^{\alpha}\>
        \sqrt{\omega_n \omega_m}}{(\omega_{1,0}^{\alpha}\!
        -\omega_n)(\omega_{2,1}^{\alpha}\!-\omega_n)}
      \\
      &&+\frac{\omega_\mathrm{an}^{\alpha}\>
        \sqrt{\omega_n \omega_m}}{(\omega_{1,0}^{\alpha}-\omega_m)
        (\omega_{2,1}^{\alpha}-\omega_m)} \biggr\}\,
        \varphi_n^*(x^\alpha) \varphi_m(x^\alpha)\, \hat{a}_n^\dagger \hat{a}_m,
      \nonumber
\end{eqnarray}
where $\omega_\mathrm{an}^{\alpha} = \omega_{1,0}^{\alpha} -
\omega_{2,1}^{\alpha}$ is the anharmonicity of the transmon spectrum. The
dispersive shift Eq.\ \eqref{eq:Hdeph1} introduces fluctuating phases in the
qubits and their thermal averaging provides us with the dephasing rates; these
are given by the irreducible correlators of the bosonic fields
\cite{Lindblad},
\begin{equation}
      \gamma_\phi^{\alpha\alpha^\prime}
        = \frac1{\hbar^2} \int d\tau \, \langle\!\langle \hat{B}^\alpha(x^\alpha,\tau)
        \hat{B}^{\alpha^\prime}(x^{\alpha^\prime},0) \rangle\! \rangle.
\end{equation}
Performing the quantum average and going to continuous frequencies one obtains
\begin{eqnarray}
      &&\gamma_\phi^{\alpha\alpha^\prime}
        = 32\pi (\kappa^\alpha \kappa^{\alpha^\prime})^2
        \int d\omega\, N_\omega(1+N_\omega) \, \\
      &&\quad \times \Bigl[ \frac{\omega_\mathrm{an}^{\alpha}\> \omega}
        {(\omega_{1,0}^{\alpha}-\omega)
        (\omega_{2,1}^{\alpha}-\omega)} \Bigr]
        \Bigl[ \frac{\omega_\mathrm{an}^{\alpha^\prime} \> \omega}
        {(\omega_{1,0}^{\alpha^\prime}-\omega)
        (\omega_{2,1}^{\alpha^\prime}-\omega)} \Bigr],
      \nonumber
\end{eqnarray}
where $N(\omega)$ is the bosonic distribution function. Finally, for
similar qubits and a narrow bandwidth $\Delta\omega$ of the
transmission line modes, one arrives at the pure dephasing rate
\begin{equation}
   \gamma_\phi = 32\pi \kappa^4\,
   \biggl[ \frac{\omega_\mathrm{an}\> \omega_0}{(\omega_{1,0}-\omega_0)
   (\omega_{2,1}-\omega_0)} \biggr]^2 \Delta\omega \, N_{\omega_0}(1+N_{\omega_0}).
\end{equation}

\subsection{Qubit relaxation}\label{app:Rel}

A lossy transmission line enhances the relaxation time of the qubits through
the Purcell effect.  The transmission line losses can be accounted for by the
dissipator
\begin{equation}\label{eq:Dline}
   {\cal D}_\mathrm{line}[\hat{R}] = \gamma_\mathrm{line} \biggl[\sum_n \hat{a}_n
   \hat{R}\, \hat{a}_n^\dagger
   - \frac12 \bigl\{ \hat{a}_n^\dagger \hat{a}_n, \hat{R}\bigr\}\biggr]
\end{equation}
in the Lindblad equation for the joint evolution of the density matrix
$\hat{R}$ of the full system, qubits and transmission line,
\begin{equation}
   \frac{d \hat{R}}{dt} = - i\bigl[ \hat{H}, \hat{R} \bigr]
   + \sum_{\alpha = {\rm \scriptscriptstyle L,R}}
   {\cal D}_\mathrm{rel}^\alpha\bigl[\hat{R}\bigr]
   + {\cal D}_\mathrm{line}\bigl[\hat{R} \bigr],
\end{equation}
with the original Hamiltonian $\hat{H}$ given by Eq.\ \eqref{eq:ham} and
${\cal D}_\mathrm{rel}^\alpha\bigl[\hat{R}\bigr]$ is a phenomenological
dissipator for the qubit $\alpha$. We perform the unitary transformation
$\hat{R} \to \hat{{\cal R}} = \hat{U}\hat{R} \hat{U}^\dagger$, with $\hat{U}$
given in Appendix \ref{app:Int_Ham} above, that integrates out the transmission
line modes in $\hat{H}$ to lowest order. Under this action, the bosonic
operators in ${\cal D}_\mathrm{line}$ are shifted according to
\begin{eqnarray}
   &&\hat{a}_n \to \hat{{\cal A}}_n = \hat{U}\hat{a}_n \hat{U}^\dagger
   \approx \hat{a}_n - \bigl[ \hat{S}^\dagger, \hat{a}_n \bigr]
   \label{eq:a}
   \\
   && = \hat{a}_n + i\sum_{\alpha,i} q_{i,i+1}^{\alpha}
   |i\rangle_{\!\alpha} \, {}_{\alpha\!}\langle i+1|\,
   \Bigl(\frac{\omega_n}{\hbar C_r} \Bigr)^{1/2}
   \frac{\varphi_n^*(x^\alpha)}{\omega_{i+1,i}^{\alpha}- \omega_n},
   \nonumber
\end{eqnarray}
where $\varphi_n(x) = \cos(\pi n x/\ell)$ and $\varphi_n(x) = i\sin(\pi n
x/\ell)$ for even and odd integers $n$, respectively. The shift includes
transitions between qubit levels and therefore the Lindblad equation for the
reduced density matrix $\hat\rho(t) = \mbox{Tr}_\mathrm{line}[\hat{R}(t)]$
assumes an additional contribution to the qubit's relaxation due to the decay
into the transmission line. Combining Eqs.\ (\ref{eq:Dline}) and (\ref{eq:a}),
the decay rate of the qubit is enhanced by the term
\begin{equation}
       \gamma_\mathrm{Pur}^{\alpha}
        = \gamma_\mathrm{line} \sum_n \frac{\omega_n}{\hbar C_r}\,
        \frac{2|q_{1,0}^{\alpha}|^2}
        {(\omega_{1,0}^{\alpha}-\omega_n)^2}
        \equiv \sum_n \gamma_{n, \mathrm{Pur}}^{\alpha},
\end{equation}
where each partial decay rate $\gamma_{n,\mathrm{Pur}}^{\alpha}$ describes the
Purcell decay into a transmission line mode with an index $n$. Going to the
continuum limit $\sum_n \to \int \frac{\ell d\omega}{2\pi v}$, one arrives at
\begin{equation}
      \gamma_\mathrm{Pur}^{\alpha}
        =  2\, \gamma_\mathrm{line} \, (\kappa^\alpha)^2 \int d\omega \frac{\omega}
        {(\omega_{1,0}^{\alpha}-\omega)^2}.
\end{equation}


\begin{thebibliography}{99}
\bibitem{entangling} While the SWAP operation replaces product states by
product states, the iSWAP gate is entangling in general.
\end{thebibliography}
\end{document}